\documentstyle[preprint,aps,prd]{revtex}
\tightenlines

\begin{document} 
\draft

\title{\bf Some remarks on the thermal and
vacuum fluctuations of a massive
\\
scalar field}

\author{R. Medina\footnote{e-mail:
        rmedina@cpd.efei.br}}
\address{Instituto de Ci\^{e}ncias\\
         Escola Federal de Engenharia de Itajub\'{a}\\
         Av. BPS 1303, 37500-000, Itajub\'{a}, Minas Gerais, Brazil}
\date{January 2001}
\preprint{gr-qc/00?????}

\maketitle
\begin{abstract}
Thermal fluctuations
of a  massive scalar field in the Rindler wedge
have been recently
obtained. As a by product, the Minkowski vacuum fluctuations seen by a 
uniformly
accelerated observer have been determined and confronted with the
corresponding Minkowski thermal fluctuations of the same field, seen by 
an inertial
observer. Since some of the calculations of this previous work have not been
detailed on it, and they present some important subtleties , they are
explicitly done here. These subtleties have to do with the leading order
behaviour of certain parameter dependent integrals. Some of the
leading order expansions are derived using the Riemann-Lebesgue lemma.

\end{abstract}


\narrowtext

\section{Introduction}
\label{int}

There is an interesting effect that arises in the study of quantum fields when
comparing the physical observables in an inertial frame and a uniformly
accelerated one: it happens that a particle detector standing on the
accelerated frame perceives a thermal bath of particles not seen by the
inertial one \cite{Unruh}. The precise mathematical statement that describes
this kind of situation relating, among other things, the thermal bath
temperature and the constant acceleration, is known as the ``Thermalization
Theorem'' (see \cite{Takagi} and references there in). So it becomes natural
to compare quantities like the thermal fluctuations of a field seen by an
inertial frame and the vacuum fluctuations of the same field
seen by a uniformly accelerated one, when using the corresponding relation
between the temperature and the acceleration.

On a recent paper, the thermal fluctuations for a massive
scalar field in the Rindler wedge were obtained \cite{paper1}. This work
included the comparison of the Minkowski vacuum fluctuations seen by a
uniformly accelerated observer and the Minkowski thermal fluctuations seen by
an inertial one. For simplicity, the detailed calculations behind those
results were omitted there. They are the main topic of this paper. The main 
results are derived for small masses. 

In section II the thermal fluctuations in Minkowski spacetime are
considered. In section III the Minkowski vacuum fluctuations seen by a
uniformly accelerated frame are studied. Section IV comments the main physical
result derived from the comparison of both situations and also comments some of
the technicalities involved in the calculations. In the Appendix it is shown
the derivation of the series and expansions needed in the main text. This
derivation includes the use of the Riemann-Lebesgue lemma in a pair of cases.

\section{Thermal fluctuations in Minkowski spacetime}
\label{min}
It was seen in \cite{paper1} that for a scalar field $\phi(x)$, 
of mass $M$, in thermal equilibrium with a heat bath at temperature 
$T$ 
and in four dimensional Minkowski spacetime, the thermal average  
$\langle\phi ^{2}(x)\rangle$ is given by
\begin{eqnarray} 
\langle\phi ^{2}(x)\rangle= 
\frac{1}{2\pi^2}\int_{0}^{\infty} 
\frac{\left(\omega^{2}+2M \omega\right)^{1/2}} 
{e^{(M+ \omega)/T}-1}\ d \omega. 
\label{vev1} 
\end{eqnarray}
In the same reference it was also seen that for $M/T\ll 1$
this expression behaves as
\begin{eqnarray}
\langle\phi ^{2}(x)\rangle= \frac{T^2}{12}- \frac{MT}{4\pi} - 
\frac{M^2}{8\pi^2} \left( \log \frac{M}{4\pi T}+\gamma-
\frac{1}{2} \right), 
\label{chave}
\end{eqnarray}
where $\gamma$ is the Euler constant and $\log$ is the neperian logarithm.

In the following lines we will derive this last result with some detail.
Using the substitution $\omega = vT - M$, the integral in (\ref{vev1}) 
may be written as
\begin{eqnarray}
\langle\phi ^{2}(x)\rangle = \frac{T^2}{2 \pi^2} \ 
I \left( \frac{M}{T} \right)
\label{vev2}
\end{eqnarray}
where
\begin{eqnarray}
I(\alpha)= \int_{\alpha}^{\infty} \frac{v}{e^v - 1} 
\left ( 1-\frac{\alpha^2}{v^2} \right ) ^{1/2} dv.
\label{Ialpha}
\end{eqnarray}
Here we are interested in the behaviour of $I(\alpha)$ for small, 
but positive,
real values of $\alpha$. For this purpose we start considering
the Taylor series
\begin{eqnarray}
(1-z)^{1/2}= 1-\frac{z}{2} - \sum_{k=2}^{\infty} a_k z^k, \ \ 
\ |z| < 1,
\label{half}
\end{eqnarray}
where 
\begin{eqnarray}
a_k=\frac{1 \cdot 3 \cdot 5 \cdot \ldots (2k-3)}{k! \cdot 2^k} \ \
 \ (k=2,3, \ldots). 
\label{ak}
\end{eqnarray}
Using $z= \alpha^2/v^2$ in (\ref{half}) , substituing in (\ref{Ialpha}) and
integrating term by term, leads to
\begin{eqnarray}
I(\alpha) = \int_{\alpha}^{\infty} 
\frac{v}{e^v - 1} dv 
- \frac{\alpha^2}{2} \int_{\alpha}^{\infty}
\frac{1}{e^v - 1} \frac{dv}{v} -
\sum_{k=2}^{\infty} a_k \alpha^{2k} 
\int_{\alpha}^{\infty} 
\frac{1}{e^v - 1} \ \frac{dv}{v^{2k-1}}.
\label{I2series}
\end{eqnarray}
Although the series in (\ref{half}) is not valid when $z=1$ (that is, when 
$v = \alpha$), and this has been used in (\ref{I2series}), it is not difficult 
to prove that the contribution of this isolated value is null when
the series is integrated term by term.   Now, in
(\ref{I2series}) we have  that
\begin{eqnarray} 
\int_{\alpha}^{\infty} \frac{v}{e^v - 1} dv = 
\frac{\pi^2}{6} - \alpha +\frac{\alpha^2}{4} + \ldots
\label{int1}
\end{eqnarray}
\begin{eqnarray}
\int_{\alpha}^{\infty} \frac{1}{e^v - 1} \frac{dv}{v}=
\frac{1}{\alpha} + \frac{1}{2} \log \alpha + 
\frac{1}{2} (\gamma - \log (2 \pi) ) + {\ldots}
\label{int2}
\end{eqnarray}
\begin{eqnarray}
\label{int3}
\int_{\alpha}^{\infty} \frac{1}{e^v - 1} \ \frac{dv}{v^{2k-1}}=
\frac{1}{2k-1} \frac{1}{\alpha^{2k-1}} 
- \frac{1}{2(2k-2)} \frac{1}{\alpha^{2k-2}} 
+ \ldots  \ \ \ (k=2,3, \cdots ), 
\end{eqnarray}
where the ``$\ldots$'' denote higher order corrections in $\alpha$.
The details of the derivation of these three formulas are given 
in Appendix I. Now, substituing (\ref{int1}), (\ref{int2}) and 
(\ref{int3}) in (\ref{I2series}), leads to
\begin{eqnarray}
I(\alpha)= \frac{\pi^2}{6} 
-\left( \frac{3}{2} + \sum_{k=2}^{\infty} \frac{a_k}{2k-1} \right) 
\alpha 
+ \left( \frac{1}{4} + \frac{1}{4}(\log (2 \pi) - \gamma )
+\frac{1}{2} \sum_{k=2}^{\infty} 
\frac{a_k}{2k-2} -\frac{1}{4} \log \alpha \right) \alpha^2 + \nonumber \\
+ \ldots .
\label{Ialpha3}
\end{eqnarray}
In Appendix II we prove that for $a_k$ given in (\ref{ak}) the 
infinite sums appearing in (\ref{Ialpha3}) are given by
\begin{eqnarray}
\sum_{k=2}^{\infty} \frac{a_k}{2k-1} = \frac{\pi}{2} - \frac{3}{2} \ 
\ \ \ \ \mbox{and}  \ \ \ \ \ 
\sum_{k=2}^{\infty} \frac{a_k}{2k-2} = \frac{1}{2} \log 2 
-\frac{1}{4}
\label{sums1}
\end{eqnarray}
So, substituing these results in (\ref{Ialpha3}) we have that
\begin{eqnarray}
I(\alpha)= \frac{\pi^2}{6} - \frac{\pi}{2} \alpha 
- \left ( \log (\frac{\alpha}{4 \pi}) + \gamma -\frac{1}{2} \right ) 
\frac{\alpha^2}{4} 
+ \ldots . 
\label{Ialpha4}
\end{eqnarray}
Upon substitution of (\ref{Ialpha4}) in (\ref{vev2}) this leads precisely to 
equation (\ref{chave}), in leading order.

\section{Minkowski vacuum fluctuations seen from a uniformly 
accelerated frame}
\label{alg}

In this section we refer the reader to Ref. \cite{Takagi} (and references
there in), where a  detailed study of the Minkowski vacuum fluctuations,
$\langle\phi ^{2}(x)\rangle$, of a
massive  scalar 
field $\phi(x)$, as seen from a uniformly accelerated frame, has 
been
done. If $a$ is the proper acceleration of the frame, then 
\cite{paper1}
\begin{eqnarray}
\langle\phi ^{2}(x)\rangle= 
\frac{1}{\pi}\int_{0}^{\infty} F(\omega) d \omega,
\label{formal}
\end{eqnarray}
where $F(\omega)$ is given in formulas $(4 \cdot 1 \cdot 12 b)$,
$(4 \cdot 1 \cdot 13 b)$ and $(4 \cdot 5 \cdot 10)$ of 
Ref. \cite{Takagi} and
may be written as 
\begin{eqnarray}
F(\omega) = \frac{1}{2 \pi} \frac{\omega}{e^{2 \pi \omega /a} -1} \ 
d_4 (\omega) 
\label{Fd4}
\end{eqnarray}
with
\begin{eqnarray}
d_4 (\omega) = \frac{2}{(\omega /a)^2 |\Gamma (i \omega/a)|^2 }
\ \int_{M/a}^{\infty} x \{ K_{i \omega /a} (x) \} ^2 dx,
\label{d4}
\end{eqnarray}
in the case of four dimensional Minkowski spacetime. In (\ref{d4}) 
$K_{i \omega /a} (x)$ is the modified Bessel function of the second 
kind
with purely imaginary index $\mu = i \omega /a$ (it is a real function).

In \cite{paper1} it was given as a result that 
$\langle\phi ^{2}(x)\rangle$ in (\ref{formal}) behaves, for $M/a \ll 1$, as  
\begin{eqnarray} 
\langle\phi ^{2}(x)\rangle = \frac{a^2}{48 \pi^2} + 
\frac{M^2}{8\pi^2} \left(\log \frac{M}{2a}+
\gamma- \frac{1}{2}+\log(-\log (\frac{M}{2a})^2) \right) . 
\label{chave2}
\end{eqnarray}
Since the derivation of this result involves, as an intermediate step, the
correction of a result found in the literature \cite{Takagi}, we will do it
here with some detail. We have done it in two steps. 
In the first one we
obtain $d_4 (\omega)$ for small values of $M/a$ 
and using this result, together with the relations in (\ref{formal}) and 
(\ref{Fd4}), we arrive to expression (\ref{chave2}) in the second step.

\vspace{0.7cm}

\noindent {\bf i) Expression for $d_4 (\omega)$ for $M/a \ll 1$} 

\vspace{0.7cm}

Using formulas $6.521-3$ and $8.332-1$ of the Gradshteyn tables
\cite{Gradshteyn}, we have that that
\begin{eqnarray}
\int_{0}^{\infty} x \{ K_{i \nu} (x) \} ^2 dx =
\frac{1}{2} \nu^2 |\Gamma(i \nu)|^2 \ \ \ \ 
\ \mbox{($\nu$ being a real number)}.
\label{K}
\end{eqnarray}
Using this result for $\nu = \omega /a$, we may write (\ref{d4}) as
\begin{eqnarray}
d_4 (\omega) = 1 - \frac{2}{(\omega /a)^2 |\Gamma (i \omega/a)|^2 }
\ \int_{0}^{M/a} x \{ K_{i \omega /a} (x) \} ^2 dx.
\label{K1}
\end{eqnarray} 
Now we look for the leading behaviour of $K_{i \omega /a} (x)$ for 
small
values of $x$, since these are the ones that give a contribution in the 
integral in (\ref{K1}) when $M/a \ll 1$.
The modified Bessel function of the second kind $K_{\beta}(x)$ is 
defined as (see \cite{Arfken}, for example)
\begin{eqnarray}
K_{\beta}(x) = \frac{\pi}{2} \ 
\frac{e^{i \beta \pi/2}\ J_{-\beta}(ix)
-e^{-i \beta \pi/2}\ J_{\beta}(ix)}
{\sin (\beta \pi)}
\label{K2}
\end{eqnarray}
where
\begin{eqnarray}
J_{\beta}(x) = \left ( \frac{x}{2} \right )^{\beta} 
\sum_{j=0}^{\infty} \frac{(-1)^j}{j! \ \Gamma(j+1+\beta)}
\left ( \frac{x}{2} \right )^{2j}
\label{J}
\end{eqnarray}
is the Bessel function of the first kind. In (\ref{K2}) and (\ref{J}) 
$ \ \beta$ is an arbitrary complex number that, in the first case, must
be different from any integer number. In the cases of integer values of
$\beta$, $K_{\beta}(x)$ is defined as the limit situation in 
(\ref{K2}).

It is a direct consequence of (\ref{K2}) and (\ref{J}) that for small 
values of $x$, $K_{i \nu} (x)$ behaves as
\begin{eqnarray}
K_{i \nu} (x) =  |\Gamma (i \nu)| \ 
\cos (\nu \log (\frac{x}{2}) - \arg \Gamma (i \nu)) + \ldots ,
\label{K3}
\end{eqnarray}
where $\arg \Gamma(i \nu)$ is the argument of the complex function
$\Gamma (i \nu)$ and ``$\ldots$'' denotes a term that vanishes
as  $x \rightarrow 0^+$. This is a peculiar result, not specifically mentioned
in commonly cited mathematical tables like Gradshteyn's \cite{Gradshteyn} or
Abramowitz's \cite{Abramowitz}, due to the purely imaginary index: it says
that for small values of $x$ the function $K_{i \nu} (x)$ oscillates 
infinitely.
Now, after 
substituing 
(\ref{K3})  in (\ref{K1}) with $\nu = \omega /a$, direct integration 
leads to
\begin{eqnarray} 
d_4 (\omega) = 1- 
\frac{M^2}{2 \omega^2 (1+\omega^2 / a^2)}
\left \{ 2 \cos ^2 \left ( \frac{\omega}{a} \log(\frac{M}{2a})- 
\arg \Gamma(\frac{i \omega}{a}) \right ) + \right.  \nonumber \\
+ \frac{\omega}{a} 
\left. \sin \ 2 \left (  \frac{\omega}{a} \log(\frac{M}{2a})- 
\arg \Gamma(\frac{i \omega}{a}) \right )+ 
\frac{\omega^2}{a^2} \right \} + \nonumber \\
+ ( \mbox{Higher order mass corrections}).
\label{d42}
\end{eqnarray}
This last expression is quite different from the one mistakenly 
derived
in equation $(4 \cdot 5 \cdot 25)$ of Ref. \cite{Takagi}. 

\vspace{0.7cm}

\noindent {\bf ii) Expression for $\langle\phi ^{2}(x)\rangle$
for $M/a \ll 1$} 

\vspace{0.7cm}

Using equation (\ref{d42}) in (\ref{Fd4}) and then in (\ref{formal}),
 $\langle\phi ^{2}(x)\rangle$ may
be  written as 
\begin{eqnarray}
\langle\phi ^{2}(x)\rangle = \frac{a^2}{48 \pi ^2} 
-\frac{M^2}{4 \pi^2} \int_{0}^{\infty} 
\frac{  2 \cos ^2  (\sigma x - \arg \Gamma(ix) ) + 
x \sin \ 2 ( \sigma x- \arg \Gamma(ix) )+x^2 }
{(e^{2 \pi x}-1) \ x \ (1+x^2)} dx + \nonumber \\
+ (\mbox{Higher order massive corrections}), 
\label{vev3}
\end{eqnarray}
where
\begin{eqnarray}
\sigma = - \log (\frac{M}{2a})
\label{sigma}
\end{eqnarray}
and where the first term, $a^2/(48 \pi ^2)$, has been obtained as   
$a^2 \zeta(2) / (8 \pi ^4)$ ($\zeta (z)$ being the Riemann Zeta function). The
condition $M/a \ll 1$ is now equivalent to $\sigma \gg 1$. Then, to study the
behaviour of the integral in (\ref{vev3}) for big values of $\sigma$ we
proceed as follows. We start using the trigonometric identities
\begin{eqnarray*}
2 \cos ^2(\alpha - \beta) = 2 \sin ^2 \alpha + 
2 \cos ^2 \beta \cos (2 \alpha) + 
\sin ( 2 \beta) \sin (2 \alpha) \ \ \ \ \ \mbox{and} \\
\sin 2 (\alpha - \beta) = 
\cos ( 2 \beta) \sin (2 \alpha) - \sin (2 \beta) \cos (2 \alpha) \ \ \ \ \ \
\end{eqnarray*}
in equation (\ref{vev3}), with $\alpha= \sigma x$ and 
$\beta = \arg \Gamma(ix)$, arriving to 
\begin{eqnarray}
\langle\phi ^{2}(x)\rangle  & = & \frac{a^2}{48 \pi ^2} - \frac{M^2}{4 \pi ^2} 
\left \{ 2 \int_{0^+}^{\infty} 
\frac{\sin ^2 ( \sigma x)}{e^{2 \pi x}-1} \frac{dx}{x} + \right. \nonumber \\
\ \ \ & \ \ & 
+ \int_{0^+}^{\infty} \frac{2 \cos ^2 (\arg \Gamma (ix))
-x \sin (2 \arg \Gamma(ix)) + x^2} 
{(e^{2 \pi x}-1) \ x \ (1+x^2)} \cos (2 \sigma x) dx + \nonumber \\ 
\ \ \ & \ \ &
\left .+ \int_{0^+}^{\infty} \frac{x \cos  (2 \arg \Gamma (ix))
+ \sin (2 \arg \Gamma(ix))}{(e^{2 \pi x}-1) \ (1+x^2)} \
\frac{\sin (2 \sigma x)}{x} dx \right \} + \nonumber \\
\ \ \ & \ \ & + (\mbox{Higher order massive corrections}).
\label{vev4}
\end{eqnarray}
Then, for each of the integrals in (\ref{vev4}) we have that
\begin{eqnarray}
\int_{0^+}^{\infty} 
\frac{\sin ^2 ( \sigma x)}{e^{2 \pi x}-1} \frac{dx}{x} & = &
- \frac{1}{4} \log (\frac{\sigma}{\sinh \sigma}) \nonumber \\
 \ & = & \frac{1}{4} ( \sigma - \log \sigma - \log 2 + \ldots )
\label{int4}
\end{eqnarray}
\begin{eqnarray}
\int_{0^+}^{\infty} \frac{2 \cos ^2 (\arg \Gamma (ix))
-x \sin (2 \arg \Gamma(ix)) + x^2} 
{(e^{2 \pi x}-1) \ x \ (1+x^2)} \cos (2 \sigma x) dx = 
0 + \ldots
\label{int5}
\end{eqnarray}
\begin{eqnarray}
\int_{0^+}^{\infty} \frac{x \cos  (2 \arg \Gamma (ix))
+ \sin (2 \arg \Gamma(ix))}{(e^{2 \pi x}-1) \ (1+x^2)} \
\frac{\sin (2 \sigma x)}{x} dx = 
\frac{1}{2} (\gamma - \frac{1}{2}) + \ldots,
\label{int6}
\end{eqnarray}
where the ``$\ldots$'' denote a term that vanishes as 
$\sigma \rightarrow \infty$.

The exact expression for the integral in (\ref{int4}) can be found in
formula $3.951-21$ of \cite{Gradshteyn}, for example. The results
in (\ref{int5}) and (\ref{int6}) are consequences of the Riemann-Lebesgue
lemma (see Ref. \cite{Whittaker}, for example). Both results
are derived in Appendix III.

Now, substituing (\ref{int4}), (\ref{int5}) and (\ref{int6}) in 
(\ref{vev4}) (with $\sigma = - \log (\frac{M}{2a})$), we arrive to
\begin{eqnarray}
\langle\phi ^{2}(x)\rangle = \frac{a^2}{48 \pi ^2} + 
\frac{M^2}{8 \pi^2} 
\left ( \log (\frac{M}{a}) + \gamma - \frac{1}{2} + 
\log (- \log(\frac{M}{2a}) \right ) + \nonumber \\
+ (\mbox{Higher order massive corrections}),
\label{vev5}
\end{eqnarray}
which is equivalent to equation (\ref{chave2}).

\section{Discussions and Final Remarks}

The thermal fluctuations in Minkowski spacetime and the Minkowski vacuum
fluctuations seen from a uniformly accelerated frame have been treated for a
massive scalar field in four dimensions. The calculations for small masses
have been explicitly done in this work and they constitute an important basis
for the results accomplished in \cite{paper1}, namely, that an accelerated
observer (with proper acceleration $a$) does not see the Minkowski vacuum as a
Minkowski thermal bath of temperature 
\begin{eqnarray}
T = \frac{a}{2 \pi},
\label{temp}
\end{eqnarray}
if the scalar field is massive. This can be seen in equations (\ref{chave}) and
(\ref{chave2}), using the relation (\ref{temp}): both expressions only agree if
the field mass is zero. The same conclusion can be reached if the situation is
analized for large masses \cite{paper1}, although this case has not been
studied here.

The derivation of $\langle \phi^2 (x) \rangle$ in (\ref{chave2}) deserves some
technical observations that we now comment.

As mentioned in section III, the
basic equations for calculating it have been taken from Ref. \cite{Takagi}.
Indeed, the expression for $d_4 (\omega)$ in (\ref{d4}) is literally taken from
\cite{Takagi}, but the small mass approximation in $(4 \cdot 5 \cdot 25)$ of
it is mistakenly done,  leading to a  result different from the one in
(\ref{d42}). Both expressions are complicated and
the easiest way of comparing them is by
analizing their behaviour when  $\omega \rightarrow 0^+$. While equation
(\ref{d42}) gives a finite value $$1 - \frac{M^2}{a^2}
(\gamma^2-\gamma+\frac{1}{2}),$$ for $d_4(\omega)$, equation $(4 \cdot 5 \cdot
25)$ gives a divergent one of the type $$\frac{M^2}{2 \omega^2}.$$ This
difference is crucial, since the vacuum fluctuations seen from a uniformly
accelerated frame are calculated as an integral of $d_4 (\omega)$ with respect
to $\omega$ (see equations (\ref{formal}) and (\ref{Fd4})).

Another observation is the usefulness of the Riemann-Lebesgue lemma in
calculating the leading behaviour of integrals of functions that involve sines
and cosines, like the one in equation (\ref{vev3}).

Finally, the checking of equation (\ref{chave2}) was done in \cite{paper1}, by
 means of comparing it with the vacuum fluctuations of a massive scalar field
in a conical spacetime, leading to complete agreement. 

\section*{ACKNOWLEDGEMENTS}

The author is thankful to Dr. Edisom Moreira Jr. for many insights and 
useful discussions. 

\section*{APPENDIX}

\subsection*{{\bf I.} Derivation of the expansions in (\ref{int1}),
(\ref{int2}) and (\ref{int3})} 
\ \ \\
Let us consider the function
\begin{eqnarray}
\label{Fp}
F_k (\alpha) = \int_{\alpha}^{\infty} \frac{1}{e^v -1} 
\frac{dv}{v^{2k-1}} \ \ \ \ (k=0, 1, 2, \ldots ), 
\end{eqnarray}
which may be written as
\begin{eqnarray}
\label{Fp1}
F_k (\alpha) = c_k + \int_{\alpha}^{1} \frac{v}{e^v -1} 
\frac{dv}{v^{2k}}, 
\end{eqnarray}
with
\begin{eqnarray}
\label{cp}
c_k = \int_{1}^{\infty} \frac{1}{e^v -1} 
\frac{dv}{v^{2k-1}}.
\end{eqnarray}
Now, we recall the power series of the generating function of the 
Bernoulli numbers, namely \cite{Arfken}
\begin{eqnarray}
\label{bern}
\frac{v}{e^v -1} = \sum_{n=0}^{\infty} B_n \frac{v^n}{n!} =
1 - \frac{1}{2} v + \frac{1}{6} v^2 + \ldots \ \ \ \ (|v| < 2 \pi) 
\end{eqnarray}
where the $B_n$ are the Bernoulli numbers.

Substituing (\ref{bern}) in (\ref{Fp1}) and integrating term by term leads to
the following behaviour for $F_k (\alpha)$:
\begin{eqnarray} 
d_k + \frac{1}{2k-1} \frac{1}{\alpha^{2k-1}} - \frac{1}{2(2k-2)}
\frac{1}{\alpha^{2k-2}} + \frac{1}{6(2k-3)} \frac{1}{\alpha^{2k-3}} + \ldots &
\ \ \ \ \mbox{if} & \ \ \ \ k = 0,2,3,4, \ldots, \nonumber \\
\frac{1}{\alpha} + \frac{1}{2} \log \alpha + d_1 + \ldots & \ \ \ \ \mbox{if} &
\ \ \ \ k=1, \nonumber \\
\label{Fp2}
\end{eqnarray}
where $d_k$ ($k=0,1,2,3, \ldots$) is the overall constant term coming out from
the integration and the ``$\ldots$'' denote higher order terms in $\alpha$.

We now analize separately the cases $k=0$, $k=1$ and $k=2,3,4, \ldots$. 

\vspace{0.7cm}

\noindent {\bf i) Case of $k=0$}

\vspace{0.7cm}

In this case, equation(\ref{Fp2}) gives
\begin{eqnarray}
\int_{\alpha}^{\infty} \frac{v}{e^v - 1} dv
= d_0 - \alpha + \frac{1}{4} \alpha^2 + \ldots.
\label{k0}
\end{eqnarray}
To derive the value of the constant term $d_0$, we consider the limit 
$\alpha \rightarrow 0^+$ in (\ref{k0}), giving
\begin{eqnarray}
d_0 = \int_{0^+}^{\infty} \frac{v}{e^v - 1} dv =
\zeta(2) = \frac{\pi^2}{6}.
\label{d0}
\end{eqnarray}
So, the results in (\ref{d0}) and (\ref{k0}) lead to the expansion in equation 
(\ref{int1}). 

\vspace{0.7cm}

\noindent {\bf ii) Case of $k=1$}

\vspace{0.7cm}

In this case, equation (\ref{Fp2}) gives
\begin{eqnarray}
\int_{\alpha}^{\infty} \frac{1}{e^v - 1} \frac{dv}{v} =
\frac{1}{\alpha} +\frac{1}{2} \log \alpha + d_1 + \ldots.
\label{F2}
\end{eqnarray}
Finding out the numerical expression for $d_1$ is quite more complicated 
than the constant term $d_0$ in the previuos case, as we will see now.

From (\ref{F2}) we can conclude that
\begin{eqnarray}
d_1 = \lim_{\alpha \rightarrow 0^+} 
\left ( \int_{\alpha}^{\infty} \frac{1}{e^v - 1} \frac{dv}{v} 
 -\frac{1}{\alpha} - \frac{1}{2} \log \alpha  \right ),
\label{d2l}
\end{eqnarray}
which may be written as
\begin{eqnarray}
d_1 = \int_{0^+}^{1} 
\left ( \frac{1}{e^v -1} - \frac{1}{v} + \frac{1}{2} \right ) \frac{dv}{v} +
\int_{1}^{\infty} 
\left ( \frac{1}{e^v -1} - \frac{1}{v}  \right ) \frac{dv}{v} .
\label{dint}
\end{eqnarray}
In order to find $d_2$ using known formulas from mathematical tables, we
rewrite expression (\ref{dint}) in the following form:
\begin{eqnarray}
d_1 = \lim_{\mu \rightarrow 0^+} 
\left \{   \int_{0^+}^{\infty} 
\left ( \frac{1}{e^v -1} - \frac{1}{v} + \frac{1}{2} \right )
\frac{e^{-\mu v}}{v} dv -
\frac{1}{2} \int_{1}^{\infty} \frac{e^{-\mu v}}{v} dv \right \},
\label{d2l1}
\end{eqnarray}
where
\begin{eqnarray}
\int_{0^+}^{\infty} 
\left ( \frac{1}{e^v -1} - \frac{1}{v} + \frac{1}{2} \right )
\frac{e^{-\mu v}}{v} dv = \log \Gamma(\mu) - (\mu - \frac{1}{2}) \log \mu +
\mu - \frac{1}{2} \log (2 \pi)
\label{G1}
\end{eqnarray}
and
\begin{eqnarray}
\int_{1}^{\infty} \frac{e^{-\mu v}}{v} dv = - \mbox{Ei}(-\mu),
\label{G2}
\end{eqnarray}
$\mbox{Ei}(x)$ being the Exponential Integral function (see formulas $3.427-4$
and $8.211-1$ in \cite{Gradshteyn} for the results in (\ref{G1}) and
(\ref{G2}), respectively).

Now we use the expansions
\begin{eqnarray}
\log \Gamma (\mu) = - \log \mu - \gamma \mu +
\sum_{k=2}^{\infty}  \frac{(-1)^k}{k} \zeta (k) \mu^k
\label{logG}
\end{eqnarray}
and
\begin{eqnarray}
\mbox{Ei}(-\mu) = \gamma + \log \mu + \sum_{k=1}^{\infty} 
\frac{(-\mu)^k}{k \cdot k!}, \label{Ei}
\end{eqnarray}
(see \cite{Arfken} and formula $8.214-1$ in \cite{Gradshteyn}, respectively) 
in (\ref{G1}) and (\ref{G2}), leading to 
\begin{eqnarray}
\int_{0^+}^{\infty} 
\left ( \frac{1}{e^v -1} - \frac{1}{v} + \frac{1}{2} \right )
\frac{e^{-\mu v}}{v} dv -
\frac{1}{2} \int_{1}^{\infty} \frac{e^{-\mu v}}{v} dv  =
\frac{1}{2} \{ \gamma - \log (2 \pi) \} + {\cal O} ( \mu \log \mu ).
\label{dint2}
\end{eqnarray}
Then, substituing (\ref{dint2}) in (\ref{d2l1}) and taking the corresponding
limit leads to
\begin{eqnarray}
d_1 = \frac{1}{2} \{ \gamma - \log (2 \pi) \} .
\label{d2f}
\end{eqnarray}
So, the results in (\ref{d2f}) and  (\ref{F2}) lead to the expansion in
equation (\ref{int2}).

\vspace{0.7cm}

\noindent {\bf iii) Case of $k=2,3,4, \ldots$}

\vspace{0.7cm}

In this case, equation (\ref{Fp2}) may be written as
\begin{eqnarray}
\int_{\alpha}^{\infty} \frac{1}{e^v - 1} \frac{dv}{v^{2k-1}} =
\frac{1}{2k-1} \frac{1}{\alpha^{2k-1}} -
\frac{1}{2(2k-2)} \frac{1}{\alpha^{2k-2}} + \ldots,
\label{F2k}
\end{eqnarray}
which is the result already mentioned in equation (\ref{int3}). There is no
interest in determining the value of the constant term in this expansion since
it is a higher order term, of no importance in our present calculations.

\subsection*{{\bf II.} Derivation of the infinite sums in (\ref{sums1}) }

Starting from the expansion of $(1-z)^{1/2}$ in a power series (given in
(\ref{half}) ) with $z=x^2$, we can derive the relations
\begin{eqnarray}
\sum_{k=2}^{\infty} a_k \ x^{2k-2} = \frac{1-x^2/2-(1-x^2)^{1/2}}{x^2}, 
\ \ \ \
\sum_{k=2}^{\infty} a_k \ x^{2k-3} = \frac{1-x^2/2-(1-x^2)^{1/2}}{x^3},
\end{eqnarray}
which are valid for $0 < x <1$.

Integrating each relation from $0^+$ to $1^-$, we have that
\begin{eqnarray}
\sum_{k=2}^{\infty} \frac{a_k}{2k-1}  = 
\int_{0^+}^{1^-} \frac{1-x^2/2-(1-x^2)^{1/2}}{x^2} dx, \ \
\sum_{k=2}^{\infty} \frac{a_k}{2k-2}  = 
\int_{0^+}^{1^-} \frac{1-x^2/2-(1-x^2)^{1/2}}{x^3} dx .
\end{eqnarray}
After finding the primitive for each integrand we finally have that
\begin{eqnarray}
\sum_{k=2}^{\infty} \frac{a_k}{2k-1}  & = & \left.  
\left \{ \frac{(1-x^2)^{3/2}}{x} + x (1-x^2)^{1/2} + \arcsin x -
\frac{1}{2} x - \frac{1}{x} \right \}  \right |_{0^+}^{1^-} \nonumber \\
\ \ \           & = & \frac{\pi}{2} - \frac{3}{2}
\end{eqnarray}
and
\begin{eqnarray}
\sum_{k=2}^{\infty} \frac{a_k}{2k-2}  & = & \left.
\frac{1}{2} \left \{ \frac{(1-x^2)^{3/2}}{x^2} + (1-x^2)^{1/2} -
\log ( (1-x^2)^{1/2} + 1) - \frac{1}{x^2} \right \} \right |_{0^+}^{1^-}
\nonumber \\ \ \ \           & = & \frac{1}{2} \log 2 - \frac{1}{4},
\end{eqnarray}
which are the infinite sums appearing in (\ref{sums1}).  

\subsection*{{\bf III.} Application of the Riemann-Lebesgue lemma to the
integrals in  (\ref{int5}) and (\ref{int6}) }

In this section we refer the reader to Chapter 9 of Ref. \cite{Whittaker} for
further details about the Riemann-Lebesgue lemma. For our present purposes we
may formulate this lemma as saying that
\begin{eqnarray}
\lim_{\rho \rightarrow \infty} \int_{0^{+}}^{\infty} f(x) \ \cos (\rho x)
\ dx = 0,
\label{R1}
\end{eqnarray} 
for a function $f(x)$ such that 
\begin{eqnarray}
\int_{0^+}^{\infty} f(x) dx 
\label{absolute}
\end{eqnarray}
converges
absolutely. The theorem is equally valid if $\cos (\rho x)$  
in (\ref{R1}) is changed by $\sin (\rho x)$.

A corollary of the Riemann-Lebesgue lemma is that \cite{Whittaker}
\begin{eqnarray}
\lim_{\rho \rightarrow \infty} \int_{0^+}^{\infty} f(x)
\frac{\sin (\rho x)}{x} dx = \frac{\pi}{2} f(0^+),
\label{R2}
\end{eqnarray}
for the same type of function as in (\ref{R1}).
Now, let us consider the functions 
\begin{eqnarray}
f_1(x) =  \frac{2 \cos ^2 (\arg \Gamma (ix))
-x \sin (2 \arg \Gamma(ix)) + x^2} 
{(e^{2 \pi x}-1) \ x \ (1+x^2)}
\label{deff1}
\end{eqnarray}
and
\begin{eqnarray}
f_2(x) =  \frac{x \cos  (2 \arg \Gamma (ix))
+ \sin (2 \arg \Gamma(ix))}{(e^{2 \pi x}-1) \ (1+x^2)},
\label{deff2}
\end{eqnarray}
for $x \ \epsilon \ (0, \infty)$, coming as part of the integrands in the
integrals of (\ref{int5}) and (\ref{int6}). For both functions the only
possible divergence happens when $x \rightarrow 0^+$. In the next lines we
will see that such a divergence does not exist, in either case. For this
purpose we start considering the expansions  
\begin{eqnarray}
\sin (2 \arg \Gamma (ix) ) = 2 \gamma x + {\cal O} (x^3), \\
\cos (2 \arg \Gamma (ix) ) = -1 + 2 \gamma^2 x^2 + {\cal O} (x^4) \ \
\mbox{and} \\
\cos ( \arg \Gamma (ix) ) = -\gamma x + {\cal O} (x^3),
\end{eqnarray}
for $ x > 0$, which are a direct consequence of the Laurent series for $\Gamma
(z)$( see \cite{Gradshteyn}, for example) with $z=ix$. So, in (\ref{deff1}) and
(\ref{deff2}) it is possible to conclude that for $x > 0$
\begin{eqnarray}
f_1 (x) = \frac{1}{\pi} ( \gamma^2 - \gamma + \frac{1}{2}) + {\cal O} (x) \ \ \
\mbox{and} \ \ \
f_2 (x) = \frac{1}{\pi} (\gamma - \frac{1}{2}) +  {\cal O} (x),
\label{deff22}
\end{eqnarray}
meaning that both functions are finite as $x \rightarrow 0^+$.

Since $f_1(x)$ and $f_2(x)$ have no singularities and they decay exponentially
for $x \rightarrow \infty$ (see equations (\ref{deff1}) and (\ref{deff2}) ),
 the requirement of absolute convergence in (\ref{absolute}) is satisfied by
them. Therefore, the Riemann-Lebesgue
lemma (\ref{R1}) and its corollary (\ref{R2}) are valid, leading respectively
to 
\begin{eqnarray}
\lim_{\rho \rightarrow \infty} \int_{0^+}^{\infty} f_1(x) \cos (\rho
x) dx = 0,
\label{R3}
\end{eqnarray}
and 
\begin{eqnarray}
\lim_{\rho \rightarrow  \infty} \int_{0^+}^{\infty} f_2(x)
\frac{\sin (\rho x)}{x} dx = \frac{1}{2} (\gamma -\frac{1}{2}).
\label{R4}
\end{eqnarray}

These two last results are equivalent to the ones in (\ref{int5}) and
(\ref{int6}), respectively, after substituing $\rho$ by $2 \sigma$ in both
limits.

\end{document}